\documentclass[british,prl,superscriptaddress,nofootinbib,reprint]{revtex4-1}
\usepackage[T1]{fontenc}
\usepackage[latin9]{inputenc}
\usepackage{babel}
\usepackage{verbatim}
\usepackage{refstyle}
\usepackage{float}
\usepackage{mathrsfs}
\usepackage{amsmath}
\usepackage{amsthm}
\usepackage{amssymb}
\usepackage{graphicx}
\usepackage[unicode=true]
 {hyperref}

\makeatletter

\AtBeginDocument{\providecommand\Secref[1]{\ref{Sec:#1}}}
\AtBeginDocument{\providecommand\Defref[1]{\ref{Def:#1}}}
\AtBeginDocument{\providecommand\Corref[1]{\ref{Cor:#1}}}
\AtBeginDocument{\providecommand\Thmref[1]{\ref{Thm:#1}}}
\AtBeginDocument{\providecommand\Lemref[1]{\ref{Lem:#1}}}
\AtBeginDocument{\providecommand\Eqref[1]{\ref{Eq:#1}}}
\AtBeginDocument{\providecommand\Figref[1]{\ref{Fig:#1}}}
\RS@ifundefined{subsecref}
  {\newref{subsec}{name = \RSsectxt}}
  {}
\RS@ifundefined{thmref}
  {\def\RSthmtxt{theorem~}\newref{thm}{name = \RSthmtxt}}
  {}
\RS@ifundefined{lemref}
  {\def\RSlemtxt{lemma~}\newref{lem}{name = \RSlemtxt}}
  {}

\theoremstyle{plain}
\newtheorem{thm}{\protect\theoremname}
\theoremstyle{plain}
\newtheorem{lem}{\protect\lemmaname}
\theoremstyle{definition}
\newtheorem{defn}{\protect\definitionname}
\theoremstyle{remark}
\newtheorem{rem}{\protect\remarkname}
\theoremstyle{definition}
 \newtheorem{example}{\protect\examplename}
\theoremstyle{plain}
\newtheorem{cor}{\protect\corollaryname}
\theoremstyle{remark}
\newtheorem{claim}{\protect\claimname}
\theoremstyle{definition}
\newtheorem*{defn*}{\protect\definitionname}

\@ifundefined{date}{}{\date{}}
\newref{thm}{name=theorem~,Name=Theorem~,names=theorems~,Names=Theorems~}
\newref{def}{name=definition~,Name=Definition~,names=definitions~,Names=Definitions~}
\newref{cor}{name=corollary~,Name=Corollary~,names=corollaries~,Names=Corollaries~}
\newref{lem}{name=lemma~,Name=Lemma~,names=lemmas~,Names=Lemmas~}
\newref{claim}{name=claim~,Name=Claim~,names=claims~,Names=Claims~}
\newref{sec}{name=section~,Name=Section~,names=sections~,Names=Sections~}
\newref{fact}{name=fact~,Name=Fact~,names=facts~,Names=Facts~}

\usepackage{color}
\definecolor{purple}{RGB}{120,20,120}

\usepackage{hyperref}

\hypersetup{
colorlinks=true,
urlcolor=cyan,
linkcolor=black,
citecolor=blue,
}

\usepackage{colortbl}

\@ifundefined{showcaptionsetup}{}{%
 \PassOptionsToPackage{caption=false}{subfig}}
\usepackage{subfig}
\makeatother

\providecommand{\claimname}{Claim}
\providecommand{\corollaryname}{Corollary}
\providecommand{\definitionname}{Definition}
\providecommand{\examplename}{Example}
\providecommand{\lemmaname}{Lemma}
\providecommand{\remarkname}{Remark}
\providecommand{\theoremname}{Theorem}

\begin{document}
\title{All fundamental non-contextuality inequalities are unique}
\author{Kishor Bharti}
\thanks{These authors contributed equally.}
\address{Centre for Quantum Technologies, National University of Singapore}
\author{Atul Singh Arora}
\thanks{These authors contributed equally.}
\address{Université libre de Bruxelles}
\author{Leong Chuan Kwek}
\address{Centre for Quantum Technologies, National University of Singapore}
\address{MajuLab, CNRS-UNS-NUS-NTU International Joint Research Unit, Singapore
UMI 3654, Singapore}
\address{National Institute of Education, Nanyang Technological University,
Singapore 637616, Singapore}
\author{Jérémie Roland}
\address{Université libre de Bruxelles}
\date{December 17, 2019}
\begin{abstract}

Contextuality is one way of capturing the non-classicality of quantum
theory. The contextual nature of a theory is often witnessed via the
violation of non-contextuality inequalities---certain linear inequalities
involving probabilities of measurement events. Using the exclusivity
graph approach (one of the two main graph theoretic approaches for
studying contextuality), it was shown {[}PRA 88, 032104 (2013); Annals
of mathematics, 51-299 (2006){]} that a necessary and sufficient condition
for witnessing contextuality is the presence of an odd number of events
(greater than three) which are either cyclically or anti-cyclically
exclusive. Thus, the non-contextuality inequalites whose underlying
exclusivity structure is as stated, either cyclic or anti-cyclic,
are fundamental to quantum theory. We show that there is a unique
non-contextuality inequality for each non-trivial cycle and anti-cycle.
In addition to the foundational interest, we expect this to aid the
understanding of contextuality as a resource to quantum computing
and its applications to local self-testing.

\end{abstract}
\maketitle

\section{Introduction}

\textsc{Motivation.} In an attempt to conceptually understand the
departure of the predictions of quantum mechanics (QM) from that of
classical physics, the notion of contextuality was introduced. It
is one of the most general ways of capturing this divergence \citep{kochen1975problem,Abramsky2011};
the celebrated Bell non-locality can be viewed as a special case of
contextuality where the context is provided via space-like separation
of the parties involved \citep{Bell64,CSW}. More generally, a \emph{context}
is defined by a set of compatible observables viz. jointly measurable
observables.

Investigations into these fundamental questions have also reaped
practical benifits. Bell non-locality, has found many applications
in quantum key distribution \citep{ekert1991quantum}, randomness
certification \citep{colbeck2009quantum}, self-testing \citep{tsirel1987quantum,summers1987bell,popescu1992generic}
and distributed computing \citep{cleve1997substituting}, to name
a few \citep{brunner2014bell}. Recently, contextuality has also been
applied more directly to quantum key distribution \citep{JKA,Cabello_QKD},
and variants of randomness certification \citep{um2013experimental},
self-testing \citep{Bharti2018}. Further, it has been uncovered to
be the resource powering the measurement based model and a class of
fault tolerant model of quantum computation \citep{Howard2014,raussendorf2013contextuality},
among others \citep{Mansfield2018} \citep{Delfosse2015,Pashayan2015,Bermejo-Vega2017,Catani2018,Spekkens2008,kochen1975problem,Howard2014,raussendorf2013contextuality,Mansfield2018}.

\textsc{Bell non-locality/contextuality.} The idea at the heart of
this discussion can be traced back to Einstein who expressed his discomfort
with the probabilistic nature of quantum mechanics by providing a
striking argument against it \citep{EPR35} using a notion of realism
(element of physical reality) for two spatially separated experiments.
He believed that there must exist local hidden variables which, once
supplied, make QM deterministic. Such \emph{completions} are referred
to as local hidden variable models. Bell constructed a linear inequality
which is violated by QM and yet it can never be violated by any such
completion \citep{Bell64}, falsifying Einstein's belief \citep{hensen2015loophole}.
It may be said that the \emph{Bell-inequality} \emph{witnesses} the
non-locality of (any such completion of) QM.

General discussions on this topic are facilitated by correspondingly
considering general probabilistic assignments to the various observable
events. The set of probabilistic assignments, which admit a local
hidden variable description, form a convex polytope (a bounded set
whose boundaries are defined by hyperplanes). The facet-defining Bell
inequalities are the characterising hyperplanes of the aforesaid polytope.
Once formalised, this becomes a general framework for studying Bell
inequalities (which can and has been refined to facilitate computations).
This can, however, be further generalised if an underlying principle
which is correspondingly more general than that of local realism,
is used. In the Bell scenario, there was a clear role of spatial separation
and therefore there were at least two parties involved. It turns out
that one can study non-classicality even for a single indivisible
quantum system. To this end, one uses \emph{non-contextual completions}
of probabilistic assignments where the phrase non-contextual emphasises
that there is a precise value assigned to each observable by the completion.
This is because it is possible to define completions where the value
assigned depends on the context (i.e. the set of compatible observables
it is measured with), and such completions can explain the predictions
of quantum mechanics. Consequently quantum mechanics is sometimes
called contextual\footnote{It is interesting to note that non-contextual completions which don't
satisfy a property known as \emph{functional consistency} can also
explain the predictions of quantum mechanics. See \citep{peres2006quantum,Arora2018}
for details.}. This schism between non-contextual completions and quantum mechanics
is used as the underlying principle to construct frameworks to study
\emph{non-contextuality (NC) inequalities}. The Klyachko-Can-Binicio\u{g}lu-Shumovsky
(KCBS) inequality may be considered to be the simplest NC inequality
(analogous to the Bell/CHSH inequality in the spatial separation setting).
There are two principal graph theoretic frameworks for studying contextuality:
the compatibility hypergraph approach \citep{Kurzynski2012,cabello2013basic,Amaral2017,amaral2018graph}
and the exclusivity graph approach \citep{CSW}. The former uses hyper-edges
to encode the compatibility relations between the observables. The
latter, uses a slightly different physical approach and focusses on
the exclusivity of measurement events. The exclusivity relations between
these events is encoded using edges\footnote{See \Secref{Relative-Simplification-Explaine} in SM for a comparison
of these two approaches.}. We show that for \emph{any} scenario (characterised by \emph{any}
given exclusivity graph) which exhibits contextuality, there is a
\emph{unique} NC inequality, corresponding to each induced subgraphs
of the graph possessing a certain property (which in turn are separately
known to exist). Further, these NC inequalities may be seen as obvious
generalisations of the KCBS inequality.

\textbf{}\textsc{Fundamental non-contextuality inequalities.} The
key appeal of the exclusivity graph approach stems from a powerful
result in graph theory---the strong perfect graph theorem \citep{chudnovsky2006strong,cabello2013basic}.
Consider a scenario encoded by a certain exclusivity graph. The contextuality
in the scenario can be witnessed by some NC inequality and appropriately
constructed states/measurements if and only if the exclusivity graph
associated with it contains, as an induced subgraph, an \emph{odd}
cyclic graph and/or an odd anti-cyclic graph of length greater than
three. The said obvious generalisation of the KCBS inequality turns
out to be the simplest inequality which has an underlying odd cycle
as its exclusivity graph. These inequalities, together with their
analogue for the anti-cycle, may in hindsight be termed \emph{fundamental
NC inequalities} \citep{cabello2013basic}. In this Letter, we show
that each odd cyclic graph and anti-cyclic graph corresponds to a
unique fundamental NC inequality, justifying its name. Given any exclusivity
graph which can exhibit contextuality, for each cycle and anti-cycle
(odd) we can directly deduce that there is a unique inequality corresponding
to it. There may, however, be additional inequalities corresponding
to other induced sub-graphs. In the supplementary material (SM), we
demonstrate this by characterising the simplest Bell exclusivity scenario
(see \Secref{All-Bell-Inequalities} in SM). In fact, the ``other
NC inequalities'' in this case turn out to be the the familiar CHSH/Bell
inequality and a \emph{Heptagonal Bell inequality}---a new, to the
best of our knowledge, Bell inequality involving seven events (see
\Secref{All-Bell-Inequalities} in SM).

\textbf{}\textsc{Relation to prior work.} In the compatibility hypergraph
approach, scenarios captured by odd $n$-cycle graphs are characterised
by $2^{n}-1$ non-trivial NC inequalities which includes the generalised
KCBS inequality \citep{Araujo2013}. In the SM, we clarify why in
the exclusivity graph approach, for odd $n$-cycle graphs, we obtain
an exponential simplification---a unique NC inequality (see \Secref{Relative-Simplification-Explaine}).
The relevance of anti-cycles is not clear in the compatibility hypergraph
approach and therefore, to the best of our knowledge, they have not
been studied. However, in the exclusivity graph approach, an easy
characterisation of anti-cycles allows us to make a much more general
statement about all scenarios (due to the strong perfect graph theorem,
as was noted).

The study of the simplest Bell scenario using the exclusivity graph
approach was carried out in \citep{Sadiq} and the fundamental NC
inequality was shown to be a Bell inequality involving only five events,
termed \emph{a Pentagonal Bell-inequality}, (while the CHSH/Bell inequality
involves eight events). However, the one involving seven events was
missing.

\section{Preliminaries}

\global\long\def\NC{\text{NC}}%
\global\long\def\Q{\text{Q}}%
\global\long\def\E{\text{E}}%
\global\long\def\abl{\text{abl }}%
\global\long\def\stab{\text{STAB}}%
\global\long\def\qstab{\text{QSTAB}}%
\global\long\def\th{\text{TH}}%
\global\long\def\conv{\text{convex hull }}%

We summarise the exclusivity graph approach here, following the work
of \citet{amaral2018graph}, deferring a more complete discussion
to the appendix. An outcome, $a$, and its associated measurement,
$M$, are together called a \emph{measurement event} (or events for
brevity) and denoted by $(a|M)$. Two events are equivalent if their
probability of occurrence is the same for all preparations. Let $p_{j}(k)$
be the probability of getting an outcome $k$ given that a measurement
$j$ was performed. Two events, $e_{i}$ and $e_{j}$ are \emph{exclusive}
if there exists a measurement $M$ such that $e_{i}$ and $e_{j}$
correspond to different outcomes of $M$ (see \Defref{exclusiveEvent}
in SM). With a family of events $\{e_{1},e_{2}\dots e_{n}\}$ we associate
the \emph{exclusivity graph}, $G:=(V,E)$ where $V$ is the set of
vertices and $E$ that of edges, whose vertices are the events and
there is  an edge between the vertices if and only if the events are
exclusive (see \Defref{exclusivityGraph} in SM). The probabilities
assigned to these events are formally given by a \emph{behaviour}
which for $G$ is defined to be a map $p:V\to[0,1]$ that assigns
to each vertex $i$ a probability $p(i)$ such that $p(i)+p(j)\le1$
for all vertices that share an edge. The map $p$ can also be seen
as a vector in $\mathbb{R}^{\left|V\right|}$ (see \Defref{behaviour}
in SM). Behaviours which admit a non-contextual completion, i.e. there
exists a non-contextual hidden variable assignment such that if the
hidden variable is traced out we recover the given behaviour, are
defined to be \emph{non-contextual behaviours} (see \Defref{nonContextualBehaviour}
in SM). The set of such behaviours is denoted by $B_{\NC}(G)$. We
can similarly define the set of \emph{quantum behaviours}, $B_{\Q}$,
to be the those which can be obtained by at least one quantum state
and corresponding observables (see \Defref{quantumBehaviour} in SM).
The set of E-principle behaviours, $B_{\E}(G)$, is one where the
behaviours respect the exclusivity principle (also referred to as
the E-principle), i.e. exclusive events must have their probability
sum to at most one (see \Defref{ePrincipleBehaviour} in SM). The
central claim of this formalism is that $B_{\NC}(G)\subseteq B_{\Q}(G)\subseteq B_{\E}(G)$
(see \Corref{centrailClaim} in SM). This is a corollary of a powerful
identification of each of the sets with geometrical objects studied
by Lovász which we describe later. We can now define more precisely
a \emph{facet-defining NC inequality} to being a non-trivial facet
of $B_{\NC}(G)$ where the direction of the inequality is chosen to
satisfy containment in $B_{\NC}(G)$ (see \Defref{facetDefiningNonContextuality}
in SM). An $n$-cycle graph is an $n$ vertex graph where every $i^{\text{th}}$
vertex is connected to the $(i+1)^{\text{th}}$ vertex (the addition
is modulo $n$). We define the \emph{fundamental cyclic non-contextuality
(FCNC) inequality}, corresponding to the $n$-cycle graph for $n$
odd, to be
\begin{equation}
K_{n}:=\sum_{i=1}^{n}p_{i}\le\frac{n-1}{2}.\label{eq:KCBS_ineq}
\end{equation}
 We analogously define the \emph{fundamental anti-cyclic non-contextuality
(FANC) inequality }for the complement of the odd $n$-cycle graph
as

\begin{equation}
K_{\bar{n}}:=\sum_{i=1}^{n}p_{i}\le2.\label{eq:anti-cycle_ineq}
\end{equation}
We recover the original KCBS inequality \citep{KCBS} in the special
case of $n=5$, both for the cyclic as well as the anti-cyclic case.

\section{Uniqueness of fundamental cyclic non-contextuality inequalities}
\begin{thm}
Consider an odd $n$-cycle exclusivity graph. The associated FCNC
inequality is a unique facet-defining NC inequality.\label{thm:main}
\end{thm}
We will need the aforementioned powerful result connecting the behaviours
to geometrically well-studied objects.
\begin{lem}
\citep{CSW} Let $e_{1},e_{2}\dots e_{n}$ be the (exclusive) events
with an associated Exclusivity Graph $G=(V,E)$. Then, 
\begin{align*}
B_{\text{NC}}(G) & =\text{STAB}(G)\\
B_{Q}(G) & =\text{TH}(G),\\
B_{E}(G) & =\text{QSTAB}(G).
\end{align*}
\end{lem}
The definition of the incidence vector, $\vec{p}_{(k)}$, and clique
are standard (see \Secref{Lovasz-Geometry} in SM). \emph{STAB($G$)}
is defined to be the convex hull of the vectors $\vec{p}_{(k)}$ for
all stable set $k$ where $\vec{p}_{(k)}$ is the incidence vector
of the set $k$. Furthermore, \emph{QSTAB-inequalities} for a graph
$G$ is the set of inequalities given by $\sum_{i\in Q}x_{i}\le1$
for every clique $Q$ of the graph. Finally, \emph{QSTAB($G$)} is
the set of vectors $x\in\mathbb{R}^{\left|V\right|}$ such that $x_{i}\ge0$,
and the QSTAB-inequalities associated with $G$ are satisfied.

\paragraph*{\protect 
}Before we prove \Thmref{main}, note that the characterisation of
STAB($G$) was given in terms of its vertices and that of QSTAB($G$)
was in terms of its hyperplanes. The following (known) link, \Lemref{STABgAndIntegerSoln},
between these representations is key to the simplification.
\begin{lem}
\citep{Groetschel2011} STAB($G$) is the convex hull of the integer
solutions to the inequalities $x_{i}\ge0$, and STAB-inequalities
for $G$, where STAB-inequalities for a graph $G=(V,E)$ are defined
to be the set of inequalities given by $(x_{i}+x_{j})\le1$ for every
$(i,j)\in E$. \label{lem:STABgAndIntegerSoln}
\end{lem}
\begin{proof}[Proof of \Thmref{main}]
We consider a $5$-cycle graph but our techniques readily generalise
to the odd $n$ cycle case (unless stated otherwise). The QSTAB inequalities,
together with the $x_{i}\ge0$ condition, can be expressed as 
\begin{align}
0\le x_{i}\le1 & \text{ for }i=\{1,2\dots5\}\label{eq:one-term}\\
x_{i}+x_{i+1}\le1 & \text{ for }i=\{1,2\dots5\}\label{eq:two-term}
\end{align}
where $i+1$ is modulo 5. Note the STAB inequalities, together with
$x_{i}\ge0$, turn out to be exactly the same as the aforesaid for
the $5$ cycle graph. (The set STAB is a convex hull of integer solutions
of STAB inequalities.) Each inequality is characterised by a hyperplane.
The vertices must lie on the intersection of (at least) five distinct
hyperplanes. From this, we can already see that the integer solutions
of STAB inequalities and the QSTAB inequalities are the same. The
FCNC inequality is one of the facet defining NC inequality. To see
this, it suffices to observe that there are exactly $5$ vertices
of STAB, whose corresponding behaviours saturate the said inequality
(since the space is $5$ dimensional)and remaining vertices satisfy
the same inequality. For the $5$ cycle case, the remaining argument
is trivial and we defer the proof of the $n$ cycle case to the end.

Note that, together with the aforementioned, if we can establish that
there is only one non-integer solution of QSTAB inequalities then
we have proven our result.

To this end, observe that there can only be the following three types
of solutions: (1) all $x_{i}$ are integers, (2) none of the $x_{i}$
are integers or $(3)$ neither all $x_{i}$ are integers nor all $x_{i}$
are non-integers (viz. at least one integer and at least one non-integer
solution).

We are interested in the latter two cases. In case 2, we can't use
any of the QSTAB inequalities involving only one term (\Eqref{one-term}).
This is because for a vertex, we saturate five distinct inequalities.
In this case, saturation of any of these inequalities will yield integer
solutions which we are not considering. Hence, the only possibility
is to use \Eqref{two-term}. Now we show that the solution is unique.
Let $x_{1}=q$ for any $0<q<1$. Saturating, we deduce $x_{2}=1-q$,
$x_{3}=q$, $x_{4}=1-q$, $x_{5}=q$ and finally $x_{1}=1-q$. This
entails $x_{1}=1-q=q$ which means $q=\text{1/2}$ uniquely.

To complete the argument, we must show that there are no solutions
in case 3. We already ruled out considering all five one term inequalities
(\Eqref{one-term}) as they yield integer solutions. Let us consider
$k$ two term inequalities (\Eqref{two-term}) and $m$ one term inequalities
such that $m+k=5$. The $m$ one term inequalities, when saturated
(because we consider the intersection of hyperplanes to obtain the
vertices), will force the corresponding $x_{i}$s to be integers.
This means that there are at least $m$ integer $x_{i}$s. To analyse
further, we consider the following game. Consider the $5$-cycle graph
(see \Figref{case3image}). Select $m$ vertices of the graph (not
to be confused with the vertices of QSTAB) and $k$ edges. The vertices
correspond to the variables fixed by the one-term inequalities (saturated,
so equalities). The edges correspond to the two-term inequalities
(again, saturated so equalities). Two cases can arise in such an assignment.
Either each of the $k$ edges is connected to one of the $m$ vertices
(possibly via other edges, if not directly) or there is at least one
edge which is not connected to any of the $m$ vertices (again, possibly
via other edges, if not directly). These two cases are represented
by the left and right graph in \Figref{case3image}. Consider the
second case. The disconnected edge (in the sense described earlier)
will correspond to a two-term equality involving two variables which
have no other constraints. This means that the set of inequalities
chosen do not uniquely determine a solution, i.e. at least one of
the inequalities chosen is redundant. This case is therefore irrelevant.
Consider the first case now. In this case, start with any one of the
$m$ vertices. This corresponds to a one-term equality which fixes
the associated variable as an integer (as was noted earlier). Now
the edge (if there is one) connected to this vertex directly, will
fix the value of the other vertex associated with the edge to be an
integer. This reasoning can be repeatedly used to show that all the
variables involved along the edges connected to the said initial vertex
are integers. This can be repeated for every one of the $m$ vertices.
This means that all variables are assigned integer values. We have
reached a contradiction which means there are no solution of the kind
assumed by case 3.
\begin{figure}[H]
\begin{centering}
\includegraphics[width=6cm]{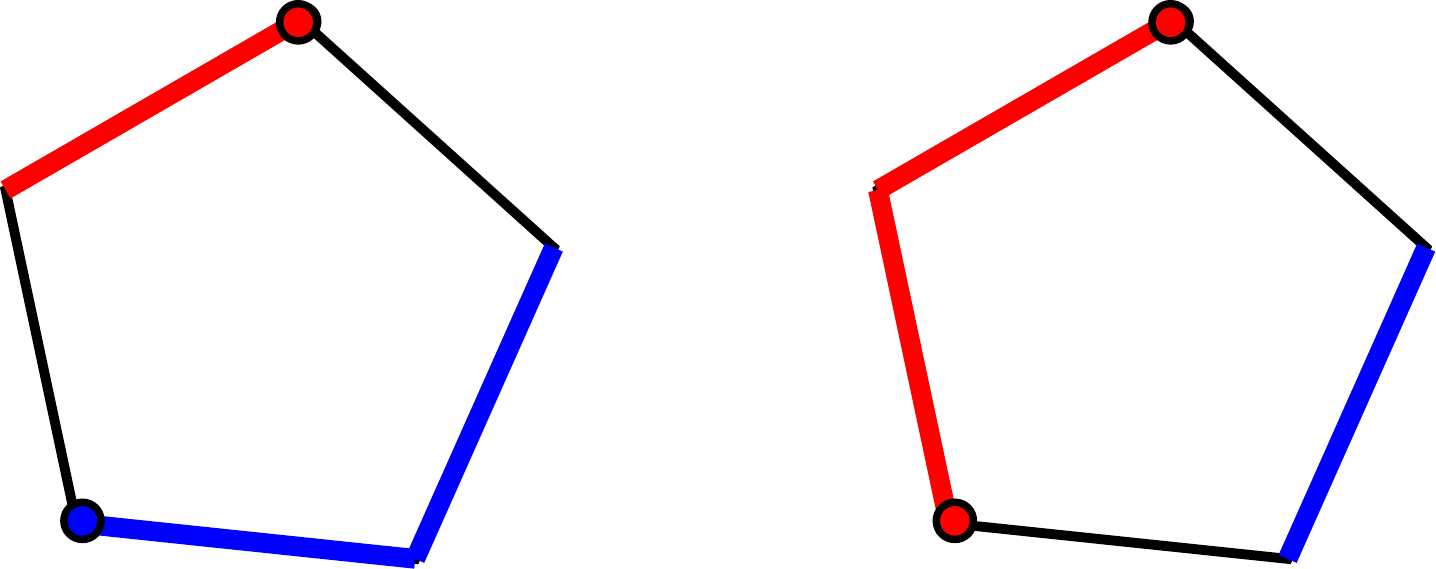}
\par\end{centering}
\caption{There are two possible scenarios corresponding to case where there
is at least one integer and one non-integer solution (case 3 in the
proof). The two-term inequalities decide the values for two $x_{i}$s
and have been represented as edges and nodes (highlighted as small
circles) have been used to denote the values determined by the one-term
inequalities. Depending on the way the combination of inequalities
is selected, one gets either all $x_{i}$s as integers or a redundant
set of inequalities leading to an undecidable value for $x_{i}$s.\label{fig:case3image}}
\end{figure}

We end by showing that the FCNC inequality is facet defining (in the
exclusivity graph approach). All incidence vectors (we will restrict
to the ones corresponding to the stable set of the $n$ cycle graph,
for this proof) will always satisfy the FCNC inequality because the
cardinality of the stable set is bounded by the independence number
(see \Defref{independenceNumber} in SM) of the graph, which for our
case is $\left(n-1\right)/2$ \citep{knuth1994sandwich,foundations2009,Liang}.
We will now show that there are exactly $n$ vertices of STAB, i.e.
incidence vectors which saturate the said inequality. To saturate,
the incidence vector must have $\left(n-1\right)/2$ components with
entry $1$, and the remaining $\left(n+1\right)/2$ components with
entry $0$. Note that each incidence vector satisfies the STAB inequalities,
i.e. if a given component is $1$ then its adjacent components are
necessarily $0$. One can convince themselves that any such vector,
i.e. incidence vectors that saturate the FCNC inequality, must have
two zeros adjacent (cyclically over $n$) while all other entries
are alternatively one and zero. The total number of ways of placing
two adjacent zeros, which is exactly $n$, then gives us the total
number of incidence vectors which saturate the inequality thereby
proving that the FCNC inequality is indeed facet defining.

\end{proof}

\section{Uniqueness of fundamental anti-cyclic inequalities}

We show that the odd anti-cycle admits a unique inequality. This follows
easily from the following known result. For any set of nonnegative
vectors $X$, its \emph{antiblocker} is defined as 
\[
\abl X=\left\{ y\ge0:x.y\le1\;\forall\:x\in X\right\} .
\]
 Let us denote by $\bar{G}$ the complement graph of $G$, viz. $\bar{G}=\left\{ V,\bar{E}\right\} $
if $G=\left\{ V,E\right\} $, which in particular means that if $G$
is a cyclic graph, then $\bar{G}$ is an anti-cyclic graph.
\begin{lem}
For any graph $G$ we have 
\begin{align*}
\stab(\bar{G}) & =\abl\qstab(G)\\
\th(\bar{G}) & =\abl\th(G)\\
\qstab(\bar{G}) & =\abl\stab(G).
\end{align*}
\label{lem:antiBlocker}
\end{lem}
Note that $\abl X=\abl\conv X$ because every element $y\in\abl X$
will satisfy $\left(\theta_{1}x_{1}+\dots+\theta_{k}x_{k}\right).y\le1$
where $\sum_{i=1}^{k}\theta_{i}=1$.
\begin{thm}
Let $G$ be odd $n$-cycle graph $(n>3)$and consider the exclusivity
graph scenario associated with $\bar{G}$. There is a unique facet-defining
NC inequality, given by $\sum_{i=1}^{n}p_{i}\le2$.
\end{thm}
\begin{proof}
We characterise $\qstab(\bar{G})$ and $\stab(\bar{G})$ by using
\Lemref{antiBlocker}. Let $\{v_{1},v_{2}\dots v_{n}\}$ denote the
vertices of $\stab(G)$. Each vertex $v_{i}$ corresponds to a hyperplane
constraining $\abl\stab(G)=\qstab(\bar{G})$. From the proof of \Thmref{main}
we know that $\qstab(G)$ has exactly one more vertex, call it $v_{0}$.
Corresponding to $v_{0}$ there will be exactly one extra hyperplane
constraining $\abl\qstab(G)=\stab(\bar{G})$ compared to those constraining
$\abl\stab(G)=\qstab(\bar{G})$. This hyperplane is precisely $\sum_{i=1}^{n}p_{i}\le2$
using $v_{0}=\left(1/2,1/2,\dots1/2\right)$ and the definition of
the anti-blocker.
\end{proof}

\section{Discussion and Conclusion}

We showed that all fundamental NC inequalities are unique for their
corresponding odd cycle (or anti-cycle) exclusivity scenario. This
is an exponential simplification compared to the compatibility hypergraph
scenario (see \Secref{Relative-Simplification-Explaine}). Any exclusivity
scenario witnessing contextuality will have associated with it at
least one fundamental NC inequality but there may be others; we give
an example of this using the simplest Bell exclusivity scenario (report
a new Bell inequality in the process; see \Secref{All-Bell-Inequalities}).
All polytopes associated with these scenarios can also be seen to
admit the same geometric interpretation which is discussed in \Secref{Geometric-Representation}
of SM.

\textbf{}A possible future direction would be to link our results
to the resource theory of contextuality. The simplification for cyclic
and anti-cyclic exclusivity scenarios indicates that there is a unique
way to capture the amount of contextuality, to wit, distance of the
contextual behaviour from the hyperplane corresponding to fundamental
NC inequalities. The perfect graph theorem could potentially allow
a generalisation to all exclusivity scenarios.

\section{Acknowledgement}

We are thankful to \href{https://www.mathworks.com/matlabcentral/profile/authors/869852-michael-kleder}{Michael Kleder}
for the CON2VERT and VERT2CON MATLAB packages that gave us the first
numerical glimpse of our result.

ASA and JR acknowledge the financial support from the Belgian Fonds
de la Recherche Scientifique - FNRS under grants no F.4515.16 (QUICTIME),
R.50.05.18.F (QuantAlgo). ASA further acknowledges the FNRS for support
through the grant F3/5/5--MCF/XH/FC--16749 FRIA. KB acknowledges
the CQT Graduate Scholarship. KB and LCK are grateful to the National
Research Foundation and the Ministry of Education, Singapore for financial
support.

\bibliographystyle{apsrev4-1}
\bibliography{theta_number}

\pagebreak{}

\onecolumngrid
\appendix

\section{All Bell Inequalities for the Simplest Exclusivity Graph\label{sec:All-Bell-Inequalities}}

\global\long\def\Ci{\text{{Ci}}}%
Given a list $[L${]} of integers, a graph with $n$ vertices where
every $i$-th vertex is connected to every other $(i+l)$ mod $n$-th
vertices for $l\in[L]$ is called a circulant graph $\text{Ci}_{n}[L].$
The exclusivity graph corresponding to the measurement events for
CHSH inequality is a circulant graph. In particular, it is represented
as $\Ci_{8}[1,4]$ graph. This is the simplest exclusivity graph which
can lead to Bell inequalities if one analyses the corresponding stable
set polytope \citep{Sadiq}.

\begin{figure}
\centering{}\includegraphics[width=4cm]{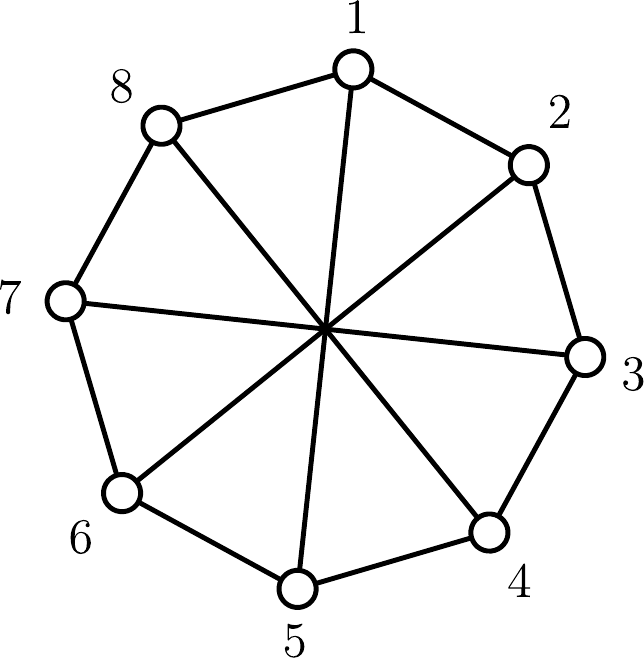}\caption{CHSH inequality is a unique non-trivial facet defining NC inequality
for $(2,2,2)$Bell scenario. The measurement events which are employed
to get the CHSH inequality follow the exclusivity relations according
to $\protect\Ci_{8}[1,4]$ graph.}
\end{figure}

We ran numerical tests to characterize the convex polytope STAB($\Ci_{8}[1,4]$)
and found three types of non-trivial facets.
\begin{enumerate}
\item Pentagonal inequalities: There are eight such inequalities and correspond
to eight induced pentagons (five cycles). In the literature, these
are known as pentagonal Bell inequalities \citep{Sadiq}. Formally,
the pentagonal inequalities are given by
\[
\sum_{i=k}^{(k+4)\text{mod}8}p_{i}\le2
\]
 for $k\in\{1,2,3,4,5,6,7,8\}.$
\item Heptagonal inequalities: There are eight such inequalities and correspond
to induced sugraphs of $\Ci_{8}[1,4]$ with any seven nodes. These
inequalities have not been reported earlier to the best of our knowledge.
Formally, the heptagonal NC inequalities are given by
\[
\sum_{i=k}^{(k+6)\text{mod}8}p_{i}\le3
\]
 for $k\in\{1,2,3,4,5,6,7,8\}.$
\item CHSH inequality: The CHSH inequality corresponds to sum of the probabilities
of all eight events and is formally given by
\[
\sum_{i=1}^{8}p_{i}\le3.
\]
\end{enumerate}
Note that neither the Heptagonal inequalities nor the CHSH inequality
correspond to any odd-cycle (or its complement).

\section{Geometric Representation\label{sec:Geometric-Representation}}

\begin{figure}[H]
\begin{centering}
\includegraphics[width=6cm]{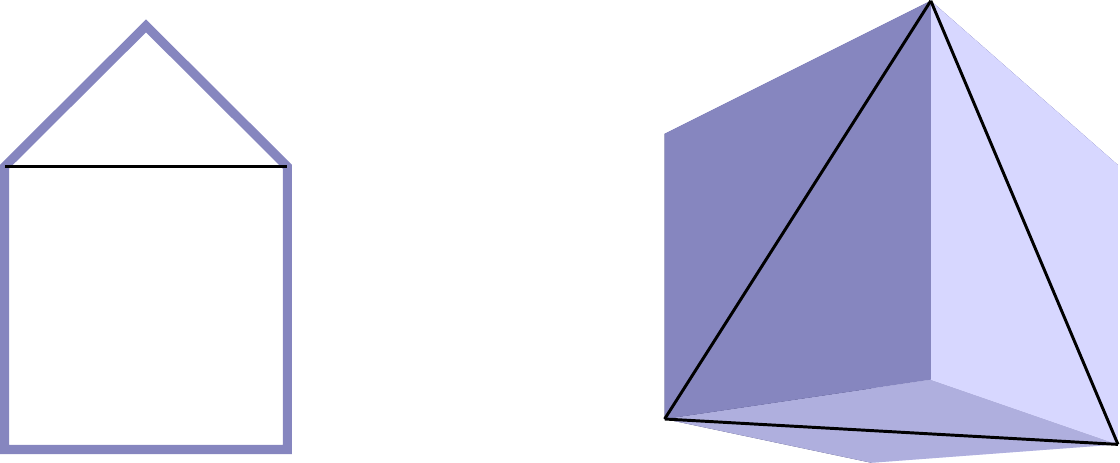}
\par\end{centering}
\caption{Two conceivable illustrations of the QSTAB polytope (light blue) and
the KCBS inequality (black) separating the said polytope from the
STAB polytope. In the left image, the vertices of both polytopes are
the same (except one) but there are two facets of QSTAB which are
not a facet of STAB. One can show that all the facets of QSTAB are
also the facets of STAB. This rules out the first image. The second
image illustrates an alternative which helps us intuitively understand
the higher dimension underlying geometry.\label{fig:Two-conceivable-illustrations}}
\end{figure}

Fix an odd $n>3$. Geometrically, the FCNC inequality corresponds
to a unique hyperplane cutting through QSTAB which separates all E-principle
behaviours (see \Defref{ePrincipleBehaviour} in Supplementary Material
(SM)) uniquely into two parts, namely non-contextual and contextual.
Naïvely one might imagine the QSTAB polytope and the FCNC inequality
to geometrically be illustrated by the image on the left in \Figref{Two-conceivable-illustrations}.
However, it is not too hard to show that all the facets of QSTAB are
also the facets of STAB which means the naïve understanding is flawed.
The image on the right in \Figref{Two-conceivable-illustrations}
better illustrates the geometry of the two convex polytopes (STAB
and QSTAB). Recall that STAB for a given graph is identical to the
anti-blocker of QSTAB for the complement graph and similarly QSTAB
for a given graph is identical to the anti-blocker of STAB for the
complement graph (see \Lemref{antiBlocker} in MT). We thus recover
essentially the same geometry for the anti-cyclic exclusivity scenario.
In summary, given an $n$-cyclic (or anti-cyclic) exclusivity scenario,
the associated fundamental NC inequality separates the corresponding
E-principle behaviours into parts and uniquely singles out a vertex
corresponding to the maximally contextual behaviour.

\section{Relative Simplification Explained\label{sec:Relative-Simplification-Explaine}}

Where is this exponential simplification coming from and are we losing
something in the process? To answer these questions we connect the
inequality we obtained using the exclusivity graph approach to the
one obtained using the compatibility hypergraph approach. We briefly
introduce the compatibility hypergraph approach first (deferring a
complete description to \Secref{Compatibility-Hypergraph-Approac}).

\subsection{Compatibility Hypergraph Approach | Overview}

Instead of looking at the exclusivity of events, the compatibility
hypergraph approach is based on, as the name suggests, the compatibility
of measurements. The scenario here is defined by a hypergraph $G$,
where the vertices $V(G)$ represent the measurements and the hyperedges
$\mathcal{C}(G)$ of the graph represent the set of measurements which
are compatible (see Definitions \ref{compatibility scenario} and
\ref{compatibility hypergraph} in SM). The set of nodes/measurements
in a hypergraph constitute a \emph{context} (see Definition \ref{Context}
in SM). For each context $C\in\mathcal{C}$, a normalised probability
vector $p_{C}$, is defined which assigns probabilities to every possible
joint outcome corresponding to the measurements belonging to that
context (see Definition \ref{Context} in SM). The behaviour $p$,
in this formalism, is defined to be the concatenation of all these
probability vectors (one vector for each context; see Definition \ref{behaviour_and_behaviour_vector}
in SM). Consider a fixed measurement. The behaviours which assign
the same marginal probability distribution to the outcomes corresponding
to this measurement, regardless of which probability vector (and hence
context) was used to evaluate the marginal, are said to be \emph{non-disturbing
behaviours}. The set of such behaviours happens to be a convex polytope
and is denoted by $\mathscr{X}(G)$ (pronounced as $X$; see Definition
\ref{ND_Set} in SM). It is analogous (and reduces, under the appropriate
restrictions) to the no-signalling distributions. Consider the set
of behaviours which arise by associating a quantum observable with
each measurement such that the compatibility requirements are satisfied.
This set is called the \emph{set of quantum behaviours} $\mathscr{Q}(G)$
and it happens to be convex. Finally, consider behaviours which admit
a non-contextual completion (see Definition \ref{def: NCC} in SM).
These behaviours can be equivalently characterised as behaviours which
arise from a global joint distribution over all the observables such
that the marginals yield the various probability vectors constituting
the behaviour (see Theorem \ref{FBA_2011} in SM). These define the
\emph{set of classical behaviours} $\mathscr{C}(G)$ which, like $\mathscr{X}(G)$,
happens to be a convex polytope. The facet-defining hyperplanes of
$\mathscr{C}(G)$ yield NC inequalities.

\subsection{The KCBS inequality}

\begin{figure}
\begin{centering}
\subfloat[KCBS in the compatibility hypergraph approach\label{fig:KCBS_compatibility_hypergraph}]{\begin{centering}
\includegraphics[width=4cm]{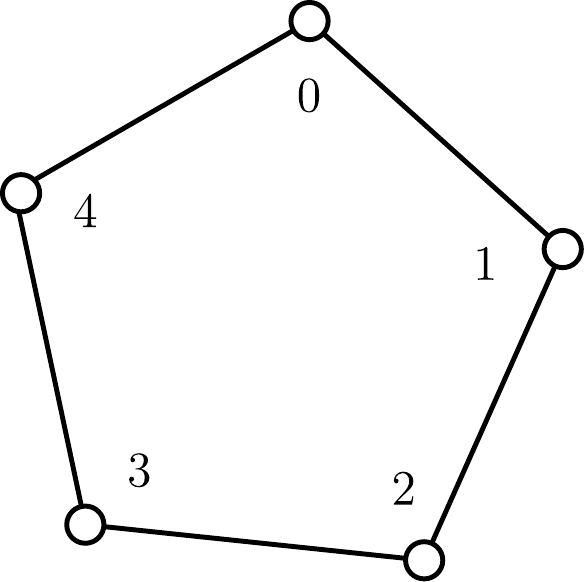}
\par\end{centering}
}\subfloat[Exclusivity graph arising from the KCBS compatibility hypergraph.
\label{fig:ExclusivityFromCompatibility}]{\begin{centering}
\includegraphics[width=8cm]{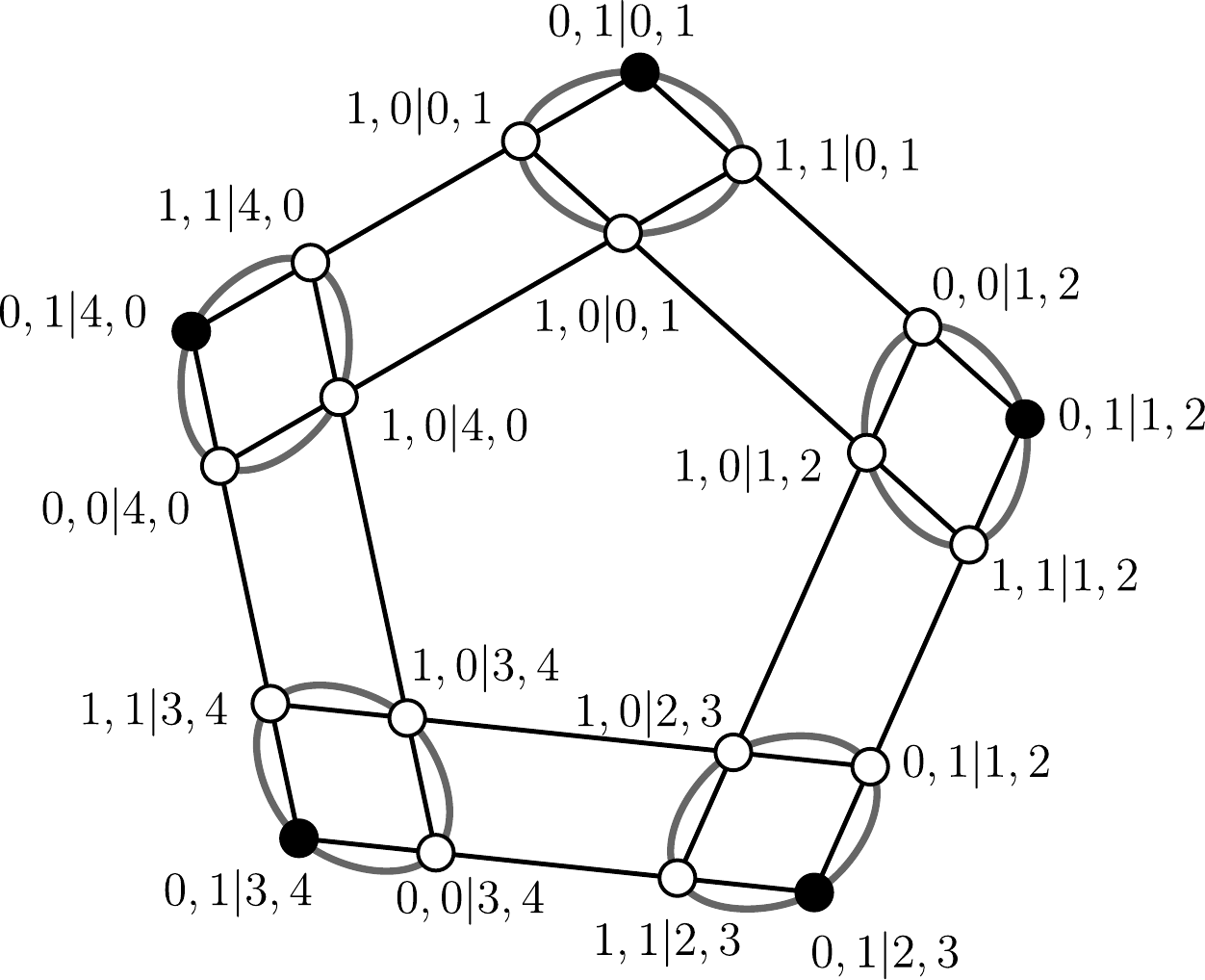}
\par\end{centering}
}
\par\end{centering}
\caption{Relation between the KCBS exclusivity and compatibility graphs}
\end{figure}

For concreteness, we consider the KCBS scenario in the compatibility
hypergraph formalism. Let $\{M_{0},M_{1},M_{2},M_{3},M_{4}\}$ be
five dichotomic measurements, which form the set of vertices $V(G)$,
and let $\{M_{0},M_{1}\},\{M_{1},M_{2}\},\{M_{2},M_{3}\},\{M_{3},M_{4}\},\{M_{4},M_{0}\}$
denote the compatibility relations among them $\mathcal{C}(G)$, which
define the (hyper)edges. This results in $G$ being a $5$-cycle (hyper)graph
(see \Figref{KCBS_compatibility_hypergraph}). It would be helpful
to use two different conventions for labelling the outcomes of the
measurements: binary $(0,1)$ and signed $(+1,-1)$. In the binary
convention, the probability of obtaining the outcome $(0,0)$ upon
the measurement of $M_{i}$ and $M_{i+1}$ is denoted by $p_{i,i+1}(00)$.
This is consistent with our notation for the probability vector $p_{C}$
where $C$ was the context, which in this case is denoted by $i,i+1$.
The addition in the indices is modulo $5$. Similarly in the signed
convention, the outcome $(+1,+1)$ is denoted by $p_{i,i+1}(++)$.
Further, to represent the expectation value of the product of measurements
$M_{i}M_{i+1}$ we use the notation 
\begin{align}
\left\langle i,i+1\right\rangle  & =p_{i,i+1}(++)+p_{i,i+1}(--)\label{eq:Exp(ij)}\\
 & \;\;-p_{i,i+1}(+-)-p_{i,i+1}(-+).\nonumber 
\end{align}
Note that we discuss expectation values only in the signed convention.
By studying the polytope of classical behaviours $\mathscr{C}(G)$
one can find the facet-defining hyperplanes which yield the NC inequalities.
These turn out to be \citep{Araujo2013}
\[
\gamma_{0}\left\langle 0,1\right\rangle +\gamma_{1}\left\langle 1,2\right\rangle +\gamma_{2}\left\langle 2,3\right\rangle +\gamma_{3}\left\langle 3,4\right\rangle +\gamma_{4}\left\langle 4,0\right\rangle \le3
\]
where $\gamma_{i}\in\{\pm1\}$ such that the number of $\gamma$s
with a negative sign are odd. To see how these relate to the unique
KCBS inequality we obtain using the exclusivity graph approach, consider
the possible events corresponding to a given context ((hyper)edge).
All these events are mutually exclusive because they correspond to
different outcomes of a given set of measurement (see \Figref{ExclusivityFromCompatibility}).
This exclusivity is denoted by the ellipse in the graph. Now consider
the measurement $M_{0}$. Let $(01|M_{0}M_{1})$ denote the event
where $M_{0}$ and $M_{1}$ are measured and the outcomes are $0$
and $1$ respectively ( see Definitions \ref{def:measurementEvent}
and \ref{def:exclusiveEvent}). Using this notation, observe that
the events $(01|M_{0}M_{1})$, $(00|M_{0}M_{1})$, $(11|M_{4}M_{0})$
and $(01|M_{4}M_{0})$ are mutually exclusive. The first two are exclusive
as they correspond to different outcomes of $M_{1}$, the second and
third are exclusive because they correspond to different outcomes
of $M_{0}$, the third and fourth are exclusive because they correspond
to different outcomes of $M_{4}$. A similar argument works for all
the other pairs. This exclusivity is denoted by a straight line. One
can verify that all the events on a given straight line are exclusive.
If we let $e_{i}=(01|M_{i}M_{i+1})$ then a 5-cycle exclusivity graph
can be extracted from the aforementioned by defining the set of vertices
to be the set of events $\{e_{0},e_{1},\dots e_{4}\}$ and the set
of edges to be their exclusivity relations $\{\left\{ e_{0},e_{1}\right\} ,\left\{ e_{1},e_{2}\right\} \dots,\left\{ e_{4},e_{0}\right\} \}$.
We had denoted the probability of these events by $p_{i}$ to write
the KCBS inequality, which in our current notation, becomes 
\begin{equation}
p_{0,1}(01)+p_{1,2}(01)+p_{2,3}(01)+p_{3,4}(01)+p_{4,0}(01)\le2.\label{eq:exclusivityKCBSinCompatibilityLikenotation}
\end{equation}
Since the exclusivity graph approach doesn't require the explicit
specification of the measurements which lead to exclusivity, merely
their existence, it is able to extract the essential nature of the
problem without creating redundancies caused by the labelling. If
the complete exclusivity graph was used then the exclusivity graph
formalism should yield effectively\footnote{related by a linear transformation}
the same NC inequalities as the compatibility hypergraph approach,
recreating the said redundancies . We now explicitly combine the following
two NCinequalities obtained using two different exclusivity graphs,
\[
\sum_{i=0}^{4}p_{i,i+1}(01)\le2,\;\sum_{i=0}^{4}p_{i,i+1}(10)\le2,
\]
into an NC inequality obtained using the compatibility hypergraph
formalism, 
\begin{equation}
-\sum_{i=0}^{4}\left\langle i,i+1\right\rangle \le3,\label{eq:KCBScompatibilityAllMinus}
\end{equation}
which corresponds to taking all $\gamma_{i}=-1$. Using $p_{i,i+1}(00)=p_{i,i+1}(++)$
(which holds by assumption; they were just different labels for the
same outcomes) in \Eqref{Exp(ij)}, and probability conservation (probability
vectors $p_{C}$ are normalised; in a given context, the probabilities
sum to one) we deduce 
\begin{equation}
\left\langle i,i+1\right\rangle =1-2(p_{i,i+1}(01)+p_{i,i+1}(10)).\label{eq:CorrTo01}
\end{equation}
To obtain \Eqref{KCBScompatibilityAllMinus} we sum the two inequalities
\begin{align*}
\sum_{i=0}^{4}\left(p_{i,i+1}(01)+p_{i,i+1}(10)\right) & \le4\\
\iff\sum_{i=0}^{4}\underbrace{\left(2p_{i,i+1}(01)+p_{i,i+1}(10)-1\right)}_{=-\left\langle i,i+1\right\rangle } & \le8-5=3.
\end{align*}
While at first sight, it might appear that the inequality obtained
using the compatibility hypergraph is weaker as it is a linear combination
of two exclusivity graph based inequalities, this conclusion is incorrect.
This is because we can do better. We can obtain from a single compatibility
hypergraph based inequality a corresponding exclusivity graph based
inequality. We show this explicitly for $\sum_{i=0}^{4}p_{i,i+1}(01)\le2$.
The marginal $p_{i}(1)=\sum_{l=0}^{1}p_{i,j}(1l)$ for every $i,j$
belonging to a hyperedge. We start with noting that $p_{i,i+1}(10)+p_{i,i+1}(01)=p_{i}(1)-p_{i,i+1}(11)+p_{i,i+1}(01)$
but we can also write $p_{i}(1)=p_{i-1,i}(01)+p_{i-1,i}(11)$ (this
is a consequence of the no-disturbance requirement). Consequently,
using \Eqref{CorrTo01}, we have 
\begin{align*}
1-\left\langle i,i+1\right\rangle  & =2\Big(p_{i-1,i}(01)+p_{i-1,i}(11)\\
 & \;\;-p_{i,i+1}(11)+p_{i,i+1}(01)\Big)\\
\implies5-\sum_{i=0}^{4}\left\langle i,i+1\right\rangle  & =4\sum_{i=0}^{4}p_{i,i+1}(01)
\end{align*}
and substituting \Eqref{KCBScompatibilityAllMinus} we obtain $4\sum_{i=0}^{4}p_{i,i+1}(01)\le5+3$
which is just \Eqref{exclusivityKCBSinCompatibilityLikenotation}.%

{[}The NC inequalities for the compatibility hypergraph structure
corresponding to anti-cycles is not known in the literature. A numerical
investigation for the classical polytope for the same indicates the
presence of exponentially many non-trivial NC inequalities. However
in the exclusivity graph approach, the non-trivial NC inequalities
corresponding to anti-cycles are unique (see Theorem ref). The reason
for the simplification is same as the cyclic case.

\section{Probabilistic Models | States and Measurements}

The results discussed here are based on the work of \citet{amaral2018graph}.
In any experimental scenario there are two types of interventions
possible, either preparation or operation. Preparation is used in
the intuitive sense of the word, that is preparing the system in a
given state, for instance using a laser to initialise the state of
an atom. More explicitly, we make the following assumptions about
the theory.
\begin{itemize}
\item Interventions are of two types: Preparation and Operation.
\item Experiments are reproducible: For each operation, there may be several
different outcomes, each occurring with a well defined probability
for a given preparation.
\end{itemize}
\begin{defn}[State]
 \label{def:state}Two preparations are defined to be equivalent
if they give the same probability distribution for all available operations.
We will refer to the equivalence class of preparations as \emph{state}. 
\end{defn}
\begin{defn}[State space]
 The set of all states is referred to as the \emph{state space} of
the system.
\end{defn}
\begin{rem}
The state space is convex.
\end{rem}
\begin{defn}[Pure states]
All extremal points of the state space are defined to be pure states.
\end{defn}
\begin{defn}[Measurements]
 \emph{Measurements} are operations with more than one outcome.
\end{defn}
\begin{rem}
Unitary evolution is an example of an operation which is not a measurement.
\end{rem}
\begin{defn}[Probabilistic model]
 We call any mathematical description of a physical system which
provides the following, a \emph{probabilistic model}.
\begin{enumerate}
\item Objects to represent
\begin{enumerate}
\item state
\item operations
\item measurements
\end{enumerate}
\item Rule to calculate the probabilities of the possible outcomes of any
arbitrary measurement given any arbitrary state.
\end{enumerate}
\end{defn}
\begin{defn}[Probability theory]
 A \emph{probability theory} is a collection of probabilistic models.
\end{defn}
\begin{defn}[Outcome repeatable measurements]
 A measurement $j$ is defined to be an outcome repeatable measurement
if every time one performs this measurement on a system and an outcome
$k$ is obtained, a subsequent measurement of $j$ on the same system
gives the outcome $k$ again with probability one.
\end{defn}
\begin{defn}[$p_{j}(k)$]
 The probability of getting an outcome $k$ given that a measurement
$j$ has been performed will be denoted by $p_{j}(k)$.
\end{defn}
All the measurements henceforth will be assumed to be outcome-repeatable.
\begin{defn}[Compatible measurements, refinement and coarse graining]
 A set of measurements $\{j_{1},j_{2}\dots j_{n}\}$ is \emph{compatible}
if there is another measurement $j$ with outcomes $\{1,\dots m\}$
and functions $\{f_{1},f_{2}\dots f_{n}\}$ such that the possible
outcomes of $j_{s}$ is the same as $\left\{ f_{s}(1),f_{s}(2)\dots f_{s}(m)\right\} $
for each $s$ and 
\[
p_{j_{s}}(l)=\sum_{k\in f_{s}^{-1}(l)}p_{j}(k)
\]
where $j$ is called a \emph{refinement} of $\{j_{1},j_{2}\dots j_{n}\}$
and each $j_{s}$ is called a \emph{coarse graining} of $j$.

If a set of measurements is compatible it is called a \emph{set of
compatible measurements}.
\end{defn}

\subsection*{Completion of a probabilistic model}
\begin{defn}[Context]
\label{Context} A set of compatible measurements is defined to be
a context.
\end{defn}
Our objective now is to construct a general mathematical framework
which can describe the completion of a probabilistic model, i.e. give
a model which is no longer probabilistic but reduces to the same probabilistic
model if certain variables are ignored.
\begin{defn}[Completion]
\label{Completion} Consider a probabilistic model $P$ where $S$
represents the set of pure states and $X$ represents the set of measurements.
The \emph{completion} of this probabilistic model, denoted by $P'$,
consists of a set of measurements $X'$, which are in one-to-one correspondence
with $X$, and a set of pure states $S'$, which are in one-to-one
correspondence with $\Lambda\times S$ for some set $\Lambda$. $P'$
must satisfy the following requirements. For all $\rho\in S$ and
all contexts $c=\{j_{1},j_{2}\dots j_{n}\}$, $P'$ should specify
a probability distribution over $\Lambda$ given by $p(\lambda)$
and a probability distribution $p_{j_{k}}^{(\lambda,\rho,c)}:\mathbb{R}\to\{0,1\}$
for each $\lambda,\rho,c,j_{k}$ such that 
\[
p_{c}^{\rho}(i_{1},i_{2}\dots i_{m})=\sum_{\lambda\in\Lambda}p(\lambda)\prod_{k=1}^{m}p_{j_{k}}^{(\lambda,\rho,c)}(i_{k})
\]
where $p_{c}^{\rho}(i_{1},i_{2}\dots i_{m})$ is the probability assigned
by $P$ to the measurement of $j_{1},j_{2}\dots j_{n}$ (encoded in
$c$) yielding the outcomes $i_{1},i_{2}\dots i_{m}$, respectively,
for the state $\rho$.
\end{defn}
\begin{rem}
We expect the completion $P'$ to specify $S'$ as $(\lambda,\rho)$
for all $\lambda\in\Lambda$ and $\rho\in S$. Let us assume for simplicity
that $X'=X$. Now for every context $c=\{j_{1},j_{2}\dots j_{n}\}$
(i.e. set of compatible measurements from $X$) the completion $P'$
will predict with certainty the outcome of measuring any $j_{i}\in c$,
for a given $(\lambda,\rho)$. This prediction is allowed to depend
on the set $c$ itself to accommodate ``contextual completions''.
We will see later that non-contextual (and functionally consistent)
completions contradict the predictions of quantum mechanics.
\end{rem}
Let $X$ be a set of measurements. Let $\{j_{1},j_{2}\dots j_{m}\}\subset X$
be a set of compatible measurements.

\begin{defn}[Non-contextual completion]
\label{def: NCC} Let $c_{1}=\left\{ j_{1},j_{2},\dots j_{m}\right\} $,
$c_{2}=\left\{ j_{1},j_{2}',\dots j_{m}'\right\} $ be two contexts
(note that $j_{i}$ and $j_{k}'$ may not be compatible for $i,k>1$).
A completion $P'$ of a probabilistic model $P$ is called \emph{non-contextual}
if $p_{j_{1}}^{(\lambda,\rho,c_{1})}(i)=p_{j_{1}}^{(\lambda,\rho,c_{2})}(i)$
for all contexts $c_{1}$ and $c_{2}$ of the aforesaid form.
\end{defn}

\section{The Exclusivity Graph Approach}

\subsection{Formalising Scenarios}

Suppose Let $(a_{1},a_{2}\dots|M_{1},M_{2}\dots)$
\begin{defn}[Measurement event]
 We denote a \emph{measurement event} by $(a_{1}a_{2}\dots a_{n}|M_{1}M_{2}\dots M_{n})$
where $a_{i}$ is the measurement outcome associated with $M_{i}$,
and $M_{i}$ is compatible with $M_{j}$ for all $i,j\in\{1,2\dots n\}$.
Two measurement events $(a_{1}a_{2}\dots a_{n}|M_{1}M_{2}\dots M_{n})$
and $(a|M)$ are \emph{equivalent} if for all states (see \Defref{state}),
their probabilities of occurrence of these events are equal.\label{def:measurementEvent}
\end{defn}
For brevity, we will use the word event in lieu of equivalent measurement
events whenever there is no ambiguity.
\begin{defn}[Exclusive event]
 Two events $e_{i}$ and $e_{j}$ are defined to be exclusive if
there exists a measurement $M$ such that $e_{i}$ and $e_{j}$ correspond
to different outcomes of $M$, i.e. $e_{i}=(a_{i}|M)$ and $e_{j}=(a_{j}|M)$
such that $a_{i}\neq a_{j}$.\label{def:exclusiveEvent}
\end{defn}
\begin{defn}[Exclusivity graph]
 For a family of events $\{e_{1},e_{2}\dots e_{n}\}$ we associate
a simple undirected graph, $G:=(V,E)$, with vertex set $V$ and edge
set $E$ such that two vertices $i,j\in V$ share an edge if and only
if $e_{i}$ and $e_{j}$ are exclusive events. $G$ is called an \emph{exclusivity
graph}.\label{def:exclusivityGraph}
\end{defn}
\begin{defn}[Probability vector]
For a given exclusivity graph $G=(V,E)$ and a probability theory,
the \emph{probability vector} is a vector $p\in\mathbb{R}^{\left|V\right|}$
such that $p_{(i)}=\text{prob}(e_{i})$ where $\text{prob}(e_{i})$
is the probability assigned by the probability theory to the event
$e_{i}$. \\
\end{defn}
\begin{defn}[Behaviour]
A \emph{behaviour} for an exclusivity graph $G=(V,E)$ is a map $p:V\to\left[0,1\right]$
which assigns to each vertex $i\in V$ a probability $p(i)$ such
that $p(i)+p(j)\le1$, for all vertices that share an edge, i.e. $\left(i,j\right)\in E(G)$.
Due to the isomorphism between the map $p:V\to[0,1]$ and the vector
$\vec{p}\in\{0,1\}^{\left|V\right|}$ we will associate with the $i^{\text{th}}$
component of $\vec{p}$ the value $p(i)$, i.e. $\vec{p}_{(i)}=p(i)$.
(Sometimes we will even drop the vector sign.) \label{def:behaviour}
\end{defn}
\begin{rem}
We don't use $p_{(i)}=p_{M}(i)$ because $M$ is not explicitly, a
priori known so cluttering the notation doesn't help.
\end{rem}

\begin{defn}[Non-contextual behaviour]
A behaviour $p$ is called a \emph{deterministic non-contextual behaviour}
if $p:V\to\{0,1\}$, i.e. $p_{(i)}\in\{0,1\}$ for all $i$ and there
exists a non-contextual completion of the corresponding probabilistic
model $P$. The set of non-contextual behaviour is defined to be the
convex hull of deterministic non-contextual behaviours and is denoted
by $B_{\text{NC}}(G)$.\label{def:nonContextualBehaviour}
\end{defn}
\begin{rem}
Defining the behaviour this way implicitly imposes functional consistency.
This is because we require a non-contextual completion of deterministic
behaviours to start with and later take its convex combination. This
imposes the exclusivity condition at the level of the hidden variable
model which in turn is a manifestation of functional consistency.
\end{rem}
\begin{defn}[Quantum behaviour]
 A behaviour for an exclusivity graph $G$ is called a quantum behaviour
if there exists a quantum state $\rho$ and projectors $\Pi_{1},\dots\Pi_{n}$
acting on a Hilbert space $\mathcal{H}$ such that $p_{(i)}=\text{Tr}(\rho\Pi_{i})$
for all $i\in V$ and $\text{Tr}(\Pi_{i}\Pi_{j})=0$ for vertices
that share an edge, i.e. $(i,j)\in E$.

The convex set of all quantum behaviours is denoted by $B_{Q}(G)$.\label{def:quantumBehaviour}
\end{defn}
\begin{defn}[The exclusivity principle]
 Given a subset $\{e_{i}\}$ of events which are pairwise exclusive
we say that the exclusivity principle is obeyed by a probabilistic
model if $\sum_{i}\text{prob}(e_{i})\le1$ for all such subsets. We
will sometimes refer to this as the E-principle.
\end{defn}
\begin{defn}[E-principle behaviour]
 A behaviour $p$ for an exclusivity graph $G$ is said to be an
E-principle behaviour if the associated probabilistic model satisfies
the exclusivity principle, i.e. $\text{prob}(e_{i})=p_{(i)}$ satisfies
the E-principle.

The set of E-principle behaviours will be denoted by $B_{E}(G)$.\label{def:ePrincipleBehaviour}
\end{defn}
Let $e_{1},e_{2}\dots e_{n}$ denote a family of measurement events.

\begin{rem}
The set $B_{\text{NC}}(G)$ is a (convex) polytope, i.e. can be expressed
as a solution of a finite number of linear inequalities.
\end{rem}
\begin{defn}[NC inequality, facet-defining]
 Let $p$ be a behaviour and $\gamma_{i},\beta\in\mathbb{R}$. A
linear inequality, $\sum\gamma_{i}p_{(i)}\le\beta$, is called an
\emph{NC inequality} of its satisfaction is a necessary condition
for membership to the set $B_{\text{NC}}(G)$. Equivalently, to claim
non-membership in the set $B_{\text{NC}}(G)$, it is sufficient to
show a violation of the said linear inequality.

An NC inequality is called \emph{facet-defining} if it defines a non-trivial
facet of $B_{\text{NC}}(G)$.\label{def:facetDefiningNonContextuality}
\end{defn}

\subsection{Lovász Geometry\label{sec:Lovasz-Geometry}}

At the risk of causing frustration by redundancy, we state the following
for clarity.
\begin{defn}
Graph: $G=(V,E)$ defined by the set of vertices and the set of edges.
\end{defn}
\begin{defn}
Orthonormal representation w.r.t. a graph $G$ is defined as follows.
For all $i\to\left|v_{i}\right\rangle $ in $\mathbb{R}^{d}$ such
that $\left\langle v_{i}|v_{j}\right\rangle =0$ whenever $(i,j)\notin E$.
\end{defn}
\begin{defn}
For a vector $\left|v_{i}\right\rangle $ in an orthonormal representation,
the cost is defined as 
\[
c_{i}=\left|\left\langle \psi|v_{i}\right\rangle \right|^{2}
\]
where $\left|\psi\right\rangle \doteq\left(1,0,\dots,0\right)$ is
a vector in $\mathbb{R}^{d}$. \label{def:costOfVector}
\end{defn}
\begin{defn}
The theta body corresponding to a graph $G$ is defined to be 
\[
\text{TH}(G)=\left\{ p\in\mathbb{R}^{\left|V\right|}\Big|p_{(i)}=c_{i}\right\} 
\]
where $c_{i}$ is the cost (see \Defref{costOfVector}) corresponding
to $\bar{G}:=\left(V,\bar{E}\right)$.
\end{defn}
\begin{defn}
Stable set/Independent set is a subset of vertices $K\subseteq V$
such that for all $i,j\in K$ there is  no edge between $i$ and $j$,
viz. $(i,j)\notin E$.
\end{defn}
\begin{defn}
Independence number of a graph $G$ is defined to be the cardinality
of the largest independent set of $G$.\label{def:independenceNumber}
\end{defn}
\begin{defn}
Clique is a subset of vertices $K\subseteq V$ such that for all $i,j\in K$
there is  an edge between $i$ and $j$, viz. $(i,j)\in E$.
\end{defn}
\begin{defn}
Incidence vector of a set is defined to be a vector $\vec{p}$ (of
size $\left|V\right|$) for $K\subseteq V$ such that 
\[
p_{(i)}=\begin{cases}
1 & \text{if }i\in K\\
0 & \text{else}.
\end{cases}
\]
\end{defn}
\begin{example}
Consider the $5$-cycle graph $V=\left\{ 1,2,3,4,5\right\} $, $E=\{(1,2),(2,3),(3,4),(4,5),(5,1)\}$.
$K=\{1,3\}$ is an example of a stable set. $K'=\{1,2\}$ is an example
of a clique. The incident vector corresponding to $K$ is $p=(1,0,1,0,0)^{T}$.
\end{example}
\begin{defn}
STAB($G$) (not to be confused with the stable set) is defined as
the convex hull of the vectors $\vec{p}_{(k)}$ for all stable sets
$k$ where $\vec{p}_{(k)}$ is the incidence vector of the set $k$.
(Note: if $k$ were an index, $\vec{p}_{(k)}$ would refer to the
$k^{\text{th}}$ component of the vector $\vec{p}$; here $k$ is
a set).
\end{defn}
\begin{defn}
QSTAB($G$) is the set of vectors $x\in\mathbb{R}^{\left|V\right|}$
such that $x_{i}\ge0$, $\sum_{i\in Q}x_{i}\le1$ for every clique
$Q$. \label{def:QSTAB}
\end{defn}
\begin{lem}
\citep{Groetschel2011} STAB($G$) is the convex hull of the integer-solutions
to the equations $x_{i}\ge0$, $\left(x_{i}+x_{j}\right)\le1$ for
every $(i,j)\in E$, where $G=(V,E)$.\label{lem:STAB}
\end{lem}
\begin{rem}
Every set of indices which is an edge is also a clique (the other
way is not necessary, obviously). This means that the inequalities
listed in \Defref{QSTAB} (viz. $\sum_{i\in Q}x_{i}\le1$ for every
clique $Q$) contain the inequalities listed in \Lemref{STAB} (viz.
$x_{i}+x_{j}\le1$ for every $(i,j)\in E$).
\end{rem}
\begin{lem}
\citep{knuth1994sandwich} STAB(G)$\subseteq$TH(G)$\subseteq$QSTAB(G).
\end{lem}

\subsection{Impossible Completions | Linking geometry and quantum mechanics}
\begin{lem}
\citep{CSW} Let $e_{1},e_{2}\dots e_{n}$ be the (exclusive) events
associated with an Exclusivity Graph $G=(V,E)$. Then, 
\begin{align*}
B_{\text{NC}}(G) & =\text{STAB}(G),\\
B_{Q}(G) & =\text{TH}(G),\\
B_{E}(G) & =\text{QSTAB}(G).
\end{align*}
\end{lem}
\begin{cor}
For a given Exclusivity Graph $G$ we have
\[
B_{\text{NC}}(G)\subseteq B_{Q}(G)\subseteq B_{E}(G).
\]
\label{cor:centrailClaim}
\end{cor}

\section{Compatibility Hypergraph Approach\label{sec:Compatibility-Hypergraph-Approac}}

\subsection{Formalising Scenarios}
\begin{defn}[Compatibility Scenario]
 \label{compatibility scenario}A \emph{compatibility scenario} is
specified by the tuple $\Upsilon:=(X,\mathcal{C},O)$ where
\begin{itemize}
\item $O$ is a finite set.
\item $X$ is a finite set of random variables from $O$ to $\mathscr{P}(O)$.
\item $\mathcal{C}$ is a collection of subsets of X such that their union
is equal to X and the intersection of any two subsets is never equal
to one of the subsets. Each of these subsets will be called a context.
\end{itemize}
\end{defn}
It might be useful to keep the following in mind to get an intuitive
understanding. The set $X$ can be thought of as the measurements,
the set $O$ as the outcome of these measurements, and the set $\mathcal{C}$
as containing maximal contexts. {[}rephrase{]}
\begin{defn}[Compatibility Hypergraph]
 \label{compatibility hypergraph}The \emph{compatibility hypergraph}
corresponding to a scenario $\Upsilon:=(X,\mathcal{C},O)$ is a hypergraph
whose nodes are elements in $X$ and hyperedges are contexts in $\mathcal{C}$.
We denote the compatibility hypergraph for the scenario by $\mathcal{H}=(X,\mathcal{C})$.
\end{defn}
We use calligraphic letters to notationally distinguish the (hyper)graphs
associated with the compatibility hypergraph from those associated
with the exclusivity graph approach.
\begin{defn}[Compatibility Graph]
 \label{compatibility graph}The \emph{compatibility graph} for a
scenario is the $2$-skeleton of the corresponding hypergraph i.e.
$\mathcal{H}=(X,\mathcal{C})$. We will denote the compatibility graph
by $\mathcal{G}=(X,E).$ Given two elements $i$ and $j$ in $X$,
they share an edge in $E$ iff $\{i,j\}\subseteq C$ for some context
$C\in\mathcal{C}.$
\end{defn}
\begin{defn}[Measurement Event]
 \label{measurement event}A \emph{measurement event} corresponds
to a single run of an experiment where the measurements in a context
$C\in\mathcal{C}$ are jointly performed with outcomes in $O^{\left|C\right|}.$
\end{defn}
Henceforth, for notational simplicity, we use $O^{C}$ instead of
$O^{\left|C\right|}$.
\begin{defn}[Behaviour and Behaviour Vector]
\label{behaviour_and_behaviour_vector} Given a scenario $\Upsilon:=(X,\mathcal{C},O)$,
a \emph{behaviour} is a family of probability distributions defined
over $O^{C}$, defined as

\[
B=\left\{ p_{C}:O^{C}\to[0,1]\Big|\sum_{s\in O^{C}}p_{C}(s)=1,C\in\mathcal{C}\right\} .
\]
One can stack the $p_{C}(s)$ for all $s\in O^{C}$ and $C\in\mathcal{C}$
to form a column vector of probabilities for a behaviour $B$. Such
probability vectors are called \emph{behaviour vectors}.
\end{defn}
\begin{rem}
The set of possible behaviours forms a polytope with $\prod_{C\in\mathcal{C}}\left|O^{C}\right|$
nodes, where the extreme points (nodes) correspond to deterministic
points i.e. $p_{C}(s)$ equal to $0$ or 1. This can be deduced from
the convexity of probability distributions.
\end{rem}
\begin{defn}[Restriction Map]
\label{Restriction_map} Given a context $C\in\mathcal{C}$ with
outcomes in $O^{C}$ and a set $U\subset C,$ a \emph{restriction
map} $r_{U}^{C}$ is given by

\[
r_{U}^{C}:O^{C}\rightarrow O^{U},
\]

\[
s=\left(a_{i}\right)_{M_{i}\in C}\mapsto s\vert_{U}=\left(a_{i}\right)_{M_{i}\in U}.
\]
\end{defn}
\begin{defn}[Marginal Distribution (for a context)]
\label{marginal_distribution} The \emph{marginal distribution} for
a probability distribution over a context $C\in\mathcal{C}$ corresponding
to a set $U\subset C$ is defined as

\[
p_{U}^{C}:O^{U}\rightarrow[0,1],
\]

\[
p_{U}^{C}(s)=\sum_{s'\in O^{C};r_{U}^{C}(s')=s}p_{C}(s'),
\]
where $r_{U}^{C}$ is a restriction map from $O^{C}$ to $O^{U}.$
\end{defn}
\begin{defn}[Non-disturbing Set, $\mathscr{X}(\Upsilon)$]
\label{ND_Set} Given a scenario $\Upsilon:=(X,\mathcal{C},O)$,
the set of behaviours is called non-disturbing set, if for any given
behaviour and two different context $C_{1}$and $C_{2}$, we have
$p_{C_{1}\cap C_{2}}^{c_{1}}=p_{C_{1}\cap C_{2}}^{c_{2}}.$
\end{defn}
\begin{defn}[Global Section]
\label{Global_section} Given a scenario $\Upsilon:=(X,\mathcal{C},O)$,
a global section for $X$ is a probalility distribution over $O^{X},$denoted
by $p_{X}:O^{X}\rightarrow[0,1].$
\end{defn}
\begin{defn}[Global Section for a behaviour]
\label{Global_section_behaviour} Given a scenario $\Upsilon:=(X,\mathcal{C},O)$
and a behaviour $B$, a global section $p_{X}$ is called a global
section for the behaviour $B$ if 
\[
p_{C}^{X}=p_{C}\quad\forall\quad C\in\mathcal{C}.
\]
\end{defn}
\begin{defn}[Non-contextual Behaviour]
\label{NC_Behaviour} A behaviour which admits a global section is
called a non-contextual behaviour.
\end{defn}
\begin{rem}
Non-contextual completions were defined independently.
\end{rem}
\begin{thm}[Fine, Brandenburger and Abramsky, 2011]
\label{FBA_2011} Given a scenario $\Upsilon:=(X,\mathcal{C},O)$
and a behaviour $B\in\mathscr{X}(\Upsilon)$, the aforementioned behaviour
$B$ has a global section if and only if there exists a non-contextual
completion recovering $B$.
\end{thm}

\subsection{Probability Distributions and Physical Theories}

\subsubsection{Classical Realisations and Noncontextuality}
\begin{rem}
Given a scenario $\Upsilon$, the set of Classical Behaviours will
be donted by $\mathscr{C}(\Upsilon)$. Note that this notation is
distinct from the $\mathcal{C}$ which was used to define the compatibility
cover for the scenario $\Upsilon$.
\end{rem}
\begin{claim}
Given a scenario $\Upsilon$ and a behaviour $B$, following statements
are equivalent:
\end{claim}
\begin{enumerate}
\item $B$ has a global section.
\item $B$ is classical.
\end{enumerate}
\begin{rem}
There exists a non-contextual completion recovering $B$.
\end{rem}

\subsubsection{Quantum Realisation}
\begin{rem}
The set of quantum behaviours will be denoted by $\mathscr{Q}(\Upsilon)$.
\end{rem}
\begin{thm}
$\mathscr{Q}(\Upsilon)$ is a convex set.
\end{thm}
\begin{rem}
Restricting the dimension of realisation in the quantum case yields
a non-convex set.
\end{rem}

\subsubsection{Non-contextuality Inequalities}
\begin{rem}
Given a scenario $\Upsilon$ and a behaviour $B$, how can one determine
if there exists a non-contextual completion recovering $B.$ This
motivates us to define linear inequalities, violation of which (for
a behavior $B$) guarantees that \textbf{$B$} is contextual i.e there
is no non-contextual completion recovering $B$. It is important to
note that the set of non-contextual behaviors forms a polytope, which
means that characterization of the same can be given by intersection
of finitely many hyperplanes and half spaces. NC inequalities for
a scenario correspond to the facets of the classical (or non-contextual)
polytope. Note that the set of non-disturbing behaviours $\mathscr{X}(\Upsilon)$
is also a polytope.

Given a polytope, the representation in terms of Half-spaces and hyperplanes
is often called $H$-representation. The same polytope can also be
described as a convex hull of finitely many vertices of the polytope,
referred to as $V$-representation.
\end{rem}
\begin{defn}[NC inequality]
\label{NCI_Compatibility} NC inequality is a linear inequality 
\[
S:=\sum_{s\in O^{C},C\in\mathscr{C}}\gamma_{C}(s)P_{C}(s)\le b
\]
where $\gamma_{C}(s)$ and $b$ are real numbers such that the inequality
is satisfied for for the behaviours in the non-contextual polytope
$\mathscr{C}(\Upsilon)$. Often $b$ is called non-contextual hidden
variable (NCHV) bound because the non-contextual behaviors (the behaviors
in $\mathscr{C}(\Upsilon)$) respect the bound.
\end{defn}
\begin{rem}
There may exist behaviours in $\mathscr{X}(\Upsilon)$ which violates
one or many NC inequalities for the scenario $\Upsilon$. Such behaviours
are often called \emph{contextual behaviours}. An NC inequality is
called \emph{tight }if there exists a non-contextual behaviour saturating
the inequality. Furthermore, it is called \emph{facet defining} if
it corresponds to one of the facets of the non-contextual polytope.
Given a behaviour $B$, its membership in $\mathscr{C}$ is equivalent
to checking if all the facet defining NC inequalities are satisfied.
\end{rem}

\subsubsection{The geometry of the case $\mathcal{H}=\mathcal{G}$ (Compatibility
Hypergraph = Compatibility Graph)}
\begin{rem}
If we assume that every context has at most 2 measurements, then the
compatibility hypergraph for the scenario is given by its $2$-skeleton.
Furthermore, if we assume that every measurement has two outcomes,
it leads to description of classical and no-disturbing sets as familiar
polytopes from graph theory.
\end{rem}

\subsubsection*{Description of the non-disturbing Quantum and non-contextual Behaviours}
\begin{rem}
The non-disturbing set $\mathscr{X}(\Upsilon)$ lies in $\mathbb{R}^{4\left|E(G)\right|}$
because every edge (or equivalently context) corresponds to two binary
measurements.
\end{rem}
\begin{defn}[Notation]
 Fix $\{M_{i},M_{j}\}\in\mathcal{C}$.
\begin{itemize}
\item $p_{ij}(ab)$ is the probalility of outcome $a$ and $b$ for the
joint measurement of $i$ and $j$ respectively.
\item $p_{i}(a)=\sum_{b}p_{ij}(ab)$
\item $p_{j}(b)=\sum_{a}p_{ij}(ab)$
\end{itemize}
\end{defn}
\begin{claim}
$p_{ij}(ab)$ can be determined from $p_{ij}(11)$ and $p_{i}(1)$
due to the constraints on non-disturbing behaviours.
\end{claim}
\begin{defn*}
$\phi:\mathbb{R}^{4\left|E(G)\right|}\to\mathbb{R}^{\left|V(G)\right|+\left|E(G)\right|}$,
$B\mapsto q=(q_{i},q_{kj})_{i\in V(G);(k,j)\in E(G)}$ where $q_{i}=p_{i}(1)$
and $q_{ij}=p_{ij}(11)$.
\end{defn*}
\begin{rem}
To return from the $q$ space to the $B$ space, we use 
\begin{align*}
p_{ij}(10) & =q_{i}-q_{ij}\\
p_{ij}(01) & =q_{j}-q_{ij}\\
p_{ij}(00) & =1-q_{i}-q_{j}+q_{ij}.
\end{align*}
\end{rem}
\begin{rem}
The map is injective.
\end{rem}
\begin{defn}[Correlation Vector $v(S)$, Correlation Polytope]
\label{CVCP} \emph{Correlation vector} $v(S)$ is defined as follows.
Given $S\subset V(G)$, $v(S)\in\mathbb{R}^{|V|+|E|}$ is defined
as 
\begin{align*}
v(S)_{i} & =\begin{cases}
1 & i\in S\\
0 & \text{else}
\end{cases} & \forall\;i\in V\\
v(S)_{ij} & =\begin{cases}
1 & i,j\in S\\
0 & \text{else}
\end{cases} & \forall\;\{i,j\}\in E.
\end{align*}
The \emph{correlation polytope} is defined to be the convex hull of
all correlation vectors.
\end{defn}
\begin{thm}
$\phi(\mathscr{C}(\Upsilon))=\text{COR}(G)$.
\end{thm}
\begin{defn}[{{[}for completeness{]} Rooted Correlation Semimetric Polytope}]
\label{RMCET} $\text{RCMET}(G)$ of a graph $G$ is the set of vectors
\[
q=(q_{i},q_{jk})\in\mathbb{R}^{\left|V(G)\right|+\left|E(G)\right|}
\]
 such that 
\begin{align*}
q_{ij} & \ge0,\\
q_{i}-q_{ij} & \ge0,\\
1-q_{i}-q_{j}+q_{ij} & \ge0.
\end{align*}
\end{defn}
\begin{thm}
$\phi(\mathscr{X}(G))=\text{RCMET}(G)$
\end{thm}

\subsubsection{The Cut Polytope}
\begin{defn}[Cut Vector, Cut Polytope]
\label{Cut_V_Cut_P_01} Given a graph $G$ and $c\in\{0,1\}^{\left|V(G)\right|}$
the Cut Vector is $x(c)\in\mathbb{R}^{|E(G)|}$ such that 
\[
x(C)_{ij}=c_{i}\oplus c_{j}.
\]
The \emph{Cut-01 Polytope, $\text{CUT}^{01}(G)$, is the convex hull
of all cut vectors of $G$.}
\end{defn}
\begin{defn}[$\pm1$ Cut Vectors]
\label{Cut_PM} Given a graph $G$ and $c\in\{0,1\}^{\left|V(G)\right|}$
the Cut Vectors are defined as $y(c)\in\mathbb{R}^{\left|E(G)\right|}$
such that 
\[
y(c)_{ij}=c_{i}c_{j}.
\]
The \emph{Cut $\pm1$ Polytope}, $\text{CUT}^{\pm1}(G)$, is the convex
hull of all $\pm1$ cut vectors of $G$.
\end{defn}
\begin{defn}[Suspension Graph $(\nabla G)$]
\label{Sus_G} The suspension graph $\nabla G$ of $G$ is the graph
with vertex-set $V(G)\sqcup\{e\}$ and edge-set $E(G)\sqcup\{(e,i),i\in V(G)\}$.
\end{defn}
\begin{rem}
$\nabla G$ is the graph obtained by adding an extra vertex and joining
all the vertices of $G$ to it.
\end{rem}
\begin{thm}
$\text{CUT}^{01}(\nabla G)=\psi(\text{COR}(G))$ where $\psi:\mathbb{R}^{|V(G)|+|E(G)|}\to\mathbb{R}^{|V(G)|+|E(G)|}$
defined by 
\[
q\mapsto x
\]
where 
\begin{align*}
x_{ij} & =1-q_{i}-q_{j}+2q_{ij} & \{i,j\}\in E(G)\\
x_{ei} & =q_{i} & i\in V(G)
\end{align*}
\end{thm}
\begin{rem}
$x_{ei}=\left\langle j\right\rangle =q_{j}=p_{j}(1)$ where $\left\langle j\right\rangle $
is the expectation value corresponding the observable $j$. Similarly,
$x_{ij}=p_{ij}(00)+p_{ij}(11)$ which is the probability of getting
the same outcome.
\end{rem}
\begin{claim}
$\text{CUT}^{\pm1}$ and $\text{CUT}^{01}$ are related by a bijective
linear map $\alpha:\text{CUT}^{01}(G)\to\text{CUT}^{\pm}(G)$ defined
by 
\[
x\mapsto y
\]
where 
\[
y_{ij}=2x_{ij}-1.
\]
\end{claim}
\begin{rem}
If we relabel the outcomes as $0\to+1$ and $1\to-1$ then we can
write 
\[
y_{ei}=-\left\langle i\right\rangle =-\left(p_{i}(1)-p_{i}(-1)\right)
\]
and 
\begin{align*}
y_{ij} & =-\left\langle ij\right\rangle \\
 & =-\left(p_{ij}(1,1)+p_{ij}(-1,-1)-p_{ij}(-1,1)-p_{ij}(1,-1)\right).
\end{align*}
\end{rem}
Now we discuss an example.
\begin{example}[Bell Scenario]
\label{Bell_Scenario} The Bell scenario corresponds to the case
where context is generated via spacelike separation of the involved
parties. We explain the defining components of the Bell scenario $\Upsilon:=(X,\mathcal{C},O)$
hereafter. We assume the number of parties to be $n$. The set $X$
consists of various disjoint subsets $X_{1},X_{2},\cdots,X_{n}.$
The subset $X_{i}$ consists of measurement operators for party $i$.
All contexts $C\in\mathcal{C}$ are of the form $C=\left\{ M_{1},M_{2},\cdots M_{n}\right\} $
where $M_{i}\in X_{i}.$ The compatibility graph for the scenario
is a complete $n$-partite graph. The Bell scenario corresponding
to $n$ party with $m$ measurements per party where each measurement
has $o$ outcomes is often denoted as $(n,m,o).$ 
\end{example}

\end{document}